\documentclass[twoside,leqno,twocolumn]{article}

\pdfoutput=1

\usepackage[letterpaper]{geometry}
\usepackage{ltexpprt}
\usepackage{graphicx}
\usepackage{todonotes}
\usepackage[colorlinks]{hyperref}

\def\Snospace~{\S{}}
\begin{document}

\title{\Large Identifying botnet IP address clusters using  natural
  language processing techniques on honeypot command logs\thanks{Work supported by the Defense Advanced Research Projects Agency (DARPA)
and the Army Contracting Command--Aberdeen Proving Grounds (ACC-APG) under Contract Number
W911NF-18-C-0020.}}
\author{Valentino Crespi \and Wes Hardaker \and Sami Abu-El-Haija \and Aram Galstyan\thanks{USC/ISI.}}
\date{}

\maketitle

\begin{abstract}
 Computer security has been plagued by increasing formidable, dynamic,
 hard-to-detect, hard-to-predict, and hard-to-characterize hacking
 techniques. Such techniques are very often deployed in
 self-propagating worms capable of automatically infecting vulnerable
 computer systems and then building large bot networks, which are then
 used to launch coordinated attacks on designated targets. In this
 work, we investigate novel applications of Natural Language
 Processing (NLP) methods to detect and correlate {\em botnet
   behaviors} through the analysis of honeypot data. In our approach
 we take observed behaviors in shell commands issued by intruders
 during captured internet sessions and reduce them to collections of
 stochastic processes that are, in turn, processed with machine
 learning techniques to build classifiers and predictors. Our
 technique results in a new ability to cluster botnet source IP
 address even in the face of their desire to obfuscate their
 penetration attempts through rapid or random permutation techniques.
\end{abstract}

\section{Introduction}

With the establishment of botnet
technologies~\cite{dagon:07,chavoshi:17,stonegross:09} the need for
methods to understand the lethality of certain behaviors and
anticipate future moves of attackers has become of paramount
importance. The challenge for cyberdefenders is further exacerbated by
the constant invention of new hard-to-detect and hard-to-track rapidly
mutating hacking techniques.  In such
 scenarios simple string-matching log searches to identify
common threat actors is no longer a sufficient technique.

In this work we study intruder behaviors within data from a honeypot network
to discover the latent characteristics of hacking agent groups.  We
base our work on the assumption that individual worms or automated
hacking techniques that attack from different IP sources are likely to
exhibit similar, but slightly varied, modes of operation. Our
intuition is that grouping behaviors by similarity of the attack
sequence may uncover botnets or coordinated agents, allowing security
analysts to track their activity, and the evolution of their attack
methodologies.
Honeypots~\cite{provos2007virtual} as a service (HAAS)
networks are large collections of Honeypots that aggregate reports
consisting of a set of time-stamped, captured {\em sessions} of
Borne-shell or similar commands collected from intrusions of
individual hackers (human or automated bot). We collect large amounts
of real and {\em unlabeled} data (see Appendix~\ref{appendix}) and apply unsupervised NLP methods to learn statistics
and cluster reports. To the best of our knowledge the application of
such methods to the cyber domain is quite novel. Supervised NLP
methods have been successfully applied to cluster {\em tweets} in
social media~\cite{andrews:19}, under the assumption that tweets from
the same account contain stationary statistics and that different
accounts would most probably correspond to different users. Here we
make opposite assumptions insofar as large sets of bots are controlled
by the same infrastructure and their individual behaviors can be periodic or
evolve dynamically. Given the impossibility of performing automated
model evaluation we chose to design two different classes of solutions
(LDA and Deep Learning) with the idea of using one as a reference for the other.




\section{Levels of attacker behaviors}
\label{sect:levels}

Suppose that at time $t$ a hacker from source IP $ip$ succeeded in
establishing a Secure Shell (SSH)~\cite{rfc4250} connection with one
of the honeypot servers and a finite sequence of session commands
$s$ is collected.  The honeypot system captures the
record~$ip:(t,s)$, where $s$ is in turn a sequence of shell
commands. Moreover, we observed that individual hackers intrude into
multiple honeypots over time. Thus, for each hacker source IP $ip$, we
observe a sequence of sessions:
$ip:(t_1,s_1),(t_2,s_2),\ldots,(t_n,s_n)$, with $t_i<t_{i+1}$ for all
$i$. We embed sessions into finite dimensional metric spaces in order
to learn the behavior of individual hackers or of groups of ``somewhat
related'' hackers (botnets) for classification and prediction
purposes.  To be precise, we are interested in two levels of analysis
of hacking behaviors: L1) the statistics of lexical terms (commands
and identifiers) occurring in its sessions and L2) given L1, the
longitudinal statistical characteristics of session time
series. Learning L1 behaviors allows us to characterize the
fuzzy-signature shell techniques used by a hacker
(\autoref{sect:L1}). Learning L2 behaviors allows us to characterize
the types of hacker activity over time, active during specific,
potentially recurring, time periods observed in certain short
subsequences of sessions (\autoref{sect:L2}).  Given the success of
these techniques, we consider future directions and approaches in
\autoref{sec:conclusions}.

\subsection{L1: learning statistics of lexical terms.}
\label{sect:L1}

We break the L1 task into three separate stages. First, we train a
probabilistic model to capture the statistics of terms occurring in
individual sessions (or in groups of sessions generated by the same
hacker\footnote{We use terms ``bot" and ``hacker" interchangeably
  throughout the text, as our technique applies to both.}). Second, we
exploit the model to embed sessions or groups of sessions into a
finite dimensional metric space. Third, we cluster such embeddings
with respect to the metric of the embedding space. The intuition
behind this method is that two behaviors are syntactically similar if
their embeddings are close in metric, namely they fall into the same
cluster. Thus, this clustering identifies botnets on the basis of
sharing similar technical sequences of commands in their generated
sessions. For the first stage we consider two different embedding
methods. In the first, we train a Latent Dirichlet Allocation (LDA)
model (\autoref{sect:lda}). In the second, we train a Deep Learning
autoencoder and exploit the encoding component to map each session to
a vector (\autoref{sect:autoenc}). LDA and, more generally, topic
modeling, provides an ``explainable'' interpretation of the embedding
that is implicit in the Bayesian description of the documents. {\em
  Par contre}, our Deep Learning autoencoder captures the order of the
commands and allows large scale processing of millions of
sessions. Finally, we employ k-means~\cite{kmeans} and
vdgmm~\cite{Blei05variationalinference} to perform clustering.

\subsubsection{LDA pipeline.}
\label{sect:lda}

\begin{figure}
    \centering
    \includegraphics[width=0.5\textwidth]{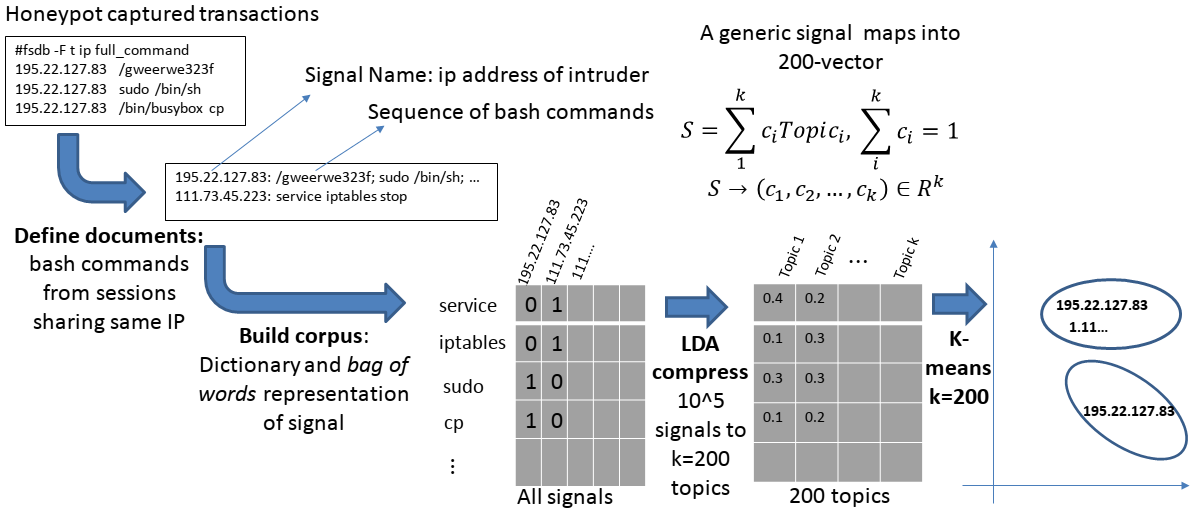}
    \caption{LDA Pipeline: ~248k documents, 200 topics.}
    \label{fig:lda}
\end{figure}

LDA is a generative probabilistic model for collections of discrete
data such as text corpora introduced in 2003 by Blei et
al.~\cite{Blei03latentdirichlet} to solve problems of text
classification. LDA models an individual document as a mixture of a
finite number of topics where, in turn, each topic is a probability
distribution over a potentially infinite vocabulary.
%
%
%
%
%
%
An LDA model consists of a collection of $K$
multinomial distributions $p(w|j)$, for $j=1,2,\ldots,K$, over ${\cal
  V}$ called {\em topics} and a $K$-dimensional Dirichlet distribution
$q(\theta)$. Given an LDA model ${\cal M}=\{p(w|j),q(\theta)\}$ a
document of $N$ words is generated in the following way. First, we
sample a multinomial distribution $\theta\sim q(\theta)$ over the $K$
topics\footnote{Samples from a $K$-dimensional Dirichlet distribution
  are probability vectors of $K$ elements that can represent a
  multinomial distribution over $K$ labels.}. Then for each
$i=1,2,\ldots,N$ we sample a topic $t_i\sim \theta(t_i)$ and the word
$w_i\sim p(w_i|t_i)$. One way of using LDA models is to apply Bayesian
techniques~\cite{Blei03latentdirichlet} to learn $p$ and $q$
from a given corpus ${\cal C}$ of documents and then, for each
document $\mathbf{w}\in {\cal C}$, to infer $\theta_{\mathbf{w}}\in
\mathbf{R}^K$ from the posterior distribution $P_{\cal
  M}(\theta|\mathbf{w})$ induced by the learned model ${\cal M}$. The
inferring of the multinomial distribution $\theta_{\mathbf{w}}$ from
${\cal M}$ and $\mathbf{w}$ defines an {\em embedding} of the document
$\mathbf{w}$ into a {\em latent} $K$-dimensional space. We process a HAAS honeypot dataset through the following pipeline
(also shown in \autoref{fig:lda}) and cluster IP addresses 
exhibiting similar statistics of lexical terms used in their sessions.

\begin{enumerate}

\item {\bf Define documents:} we aggregate sessions by source IP
  yielding a collection ${\cal D}_1$ of documents;
 
\item {\bf Build corpus:} we tokenize documents in ${\cal D}_1$,
  build a vocabulary ${\cal V}_1$, and encode each document as a
  ``bag of words'' with respect to ${\cal V}_1$. This gives a corpus
  ${\cal C}_1$ of encoded documents;
    
\item {\bf LDA compress:} we train an LDA model ({\tt gensim}) with
  corpus ${\cal C}_1$ for a fixed number of topics~\footnote{A commonly chosen value in NLP.} ($K=200$) and we
  exploit the model to embed each document in ${\cal C}_1$ in the
  latent $K$-dimensional space;

\item {\bf Cluster:} we employ $k$-means with $k=200$ to cluster all
  the embedded documents in the latent space and group IPs by
  similarity of fuzzy signature.

\end{enumerate}

\subsubsection{Autoencoder pipeline.}
\label{sect:autoenc}

We also employ a second NLP method for comparison using applications
of Deep Learning to unsupervised clustering (Deep Clustering), which
proved effective in clustering unlabeled images~\cite{guo:17,xie:16}
and signals~\cite{madiraju2018deep}. Our deep learning pipeline,
detailed below, consists of two deep learning modules. First, we
tokenize all sessions as before to build a vocabulary ${\cal V}$ and
then train a Word2Vec model~\cite{levy:14}. This defines a map
$W:{\cal V}\rightarrow \mathbf{R}^h$, $h=128$, that embeds each word
into a $h$-dimensional metric space, and, by extension, each sequence
of tokens $v_1v_2\cdots v_N$ into the $N\times h$ matrix
$W(v_1^N)=[W(v_1);W(v_2);\ldots; W(v_N)]\in\mathbf{R}^{N\times h}$
(here, we have used the expression $v_i^j$ to denote the sequence $v_i
v_{i+1} \ldots v_j$, $\mathbf{R}^{N\times h}$ to denote the set of all
$N\times h$ matrices of real numbers, and employed matlab syntax to
describe matrices). Second, for a fixed value of $N$ (e.g., $N=500$,
sessions are either truncated or padded in order to have a fixed
length in number of tokens) we train an {\em ad hoc} autoencoder
$D\circ E:\mathbf{R}^{N\times h}\rightarrow\mathbf{R}^{N\times h}$,
with $E:\mathbf{R}^{N\times h}\rightarrow \mathbf{R}^M$,
$D:\mathbf{R}^M\rightarrow\mathbf{R}^{N\times h}$, and
$D(E(W(s)))\approx W(s)$ for each (tokenized) session $s$.

\begin{figure}
    \centering
    \includegraphics[width=0.48\textwidth]{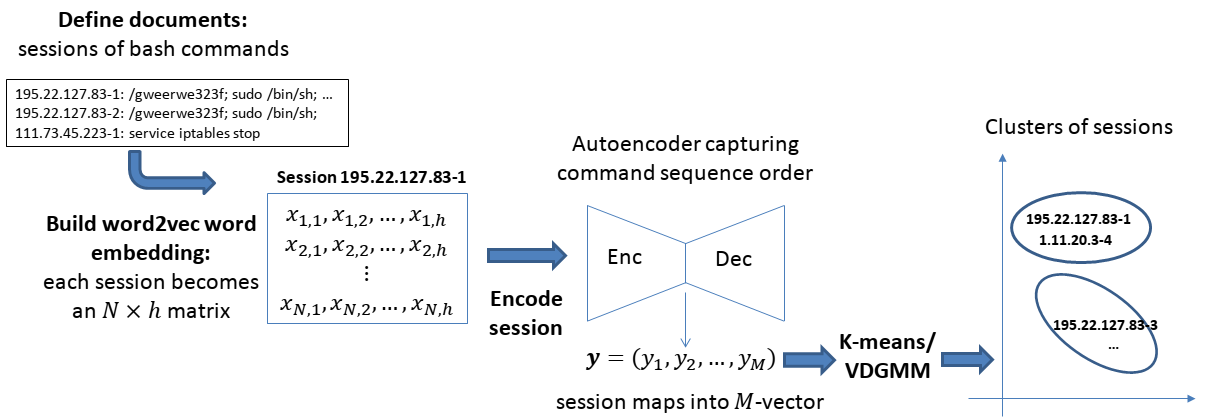}
    \caption{Autoencoder Pipeline: 420k source addresses, 90M sessions
      over 3 years, $N=500$ (length of sessions), $h=128$ (dimension
      of word embedding space), $M=200$ (dimension of session
      embedding space).}
    \label{fig:autoenc_pipeline}
\end{figure}

We implemented a standard Word2Vec {\em Skip-gram} model~\footnote{We considered two models: 1) different tokens map to different IDs; 2) tokens occurring only once map to the same ID.}
with {\em Negative Sampling} (with a window size of 2 and a number of
negative samples of 4) using tensorflow and experimentally established
that $h=128$ (with 4 epochs of training) was an optimal embedding
dimension. We designed an autoencoder to capture the
order of tokens in sessions. The Encoder consists of an upper LSTM
layer built on top of a (maxpooled) {\em convolutional} lower layer,
and the decoder consists of an upper {\em deconvolution} layer
built on top of an {\em upsampling} layer. The intuition behind this
architecture is that, while convolutional layers extract latent
features, LSTMs capture the order of the sequence of the tokens in
their latent representation. The output of the encoder
$E(W(s))\in\mathbf{R}^M$, $M=200$, provides the embedding of session
$s$ into an $M$-dimensional metric space. Unlike in~\cite{guo:17} we did not jointly
train a clustering layer together with the autoencoder because we were
interested in also learning the number of session
clusters. Thus, we clustered the embedded sessions employing
a Bayesian nonparametric algorithm (VDGMM) that learns the parameters
of a Gaussian Mixture Model (GMM) with an {\em a priori} unknown
number of Gaussian components~\cite{Blei05variationalinference}. As before, we process the HAAS data with the following
pipeline (see \autoref{fig:autoenc_pipeline}) to discover clusters of ``similar''
sessions:


\begin{enumerate}

\item {\bf Define documents:} we treat each session as a separate
  document and build collection ${\cal D}_2$ (with {\tt 192.0.2.1-4}
  we mean the $5^{\rm th}$ recorded session from {\tt 192.0.2.1});

\item {\bf Build word embedding:} we tokenize all the documents in
  ${\cal D}_2$ and build a vocabulary ${\cal V}_2$. We then train a
  Word2Vec model and exploit it to encode each document as an $N\times
  h$ matrix. This gives a corpus ${\cal C}_2$ of encoded sessions;
    
\item {\bf Build session embedding:} we use corpus ${\cal C}_2$ to train
  an autoencoder whose encoder component encodes inputs into vectors
  of $M$ reals, $M=200$. We exploit the encoder to map each session in
  ${\cal C}_2$ to a vector in $\mathbf{R}^M$;

\item {\bf Cluster:} we employ VDGMM to cluster embedded sessions
  recorded in a bounded period of time (a month in our experiments)
  and learn the number of different types of sessions.

\end{enumerate}

\subsubsection{L1 clustering results.}
\label{sect:results_lda}

\begin{figure}
    \centering
    \includegraphics[width=0.48\textwidth]{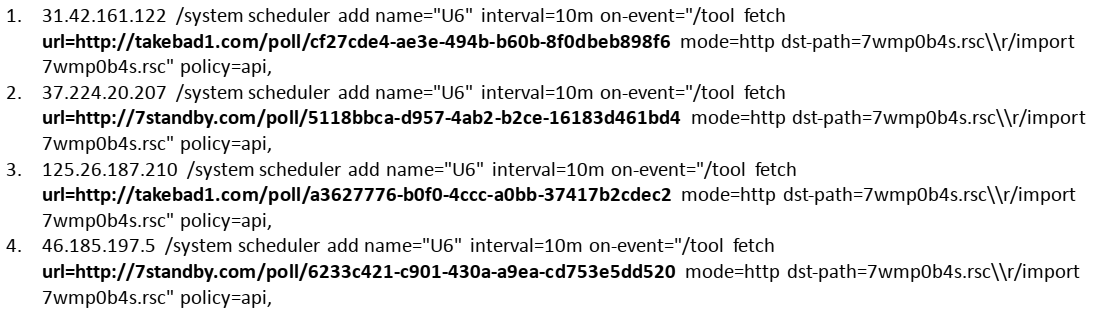}
    \caption{First $4$ documents in a random cluster from the LDA pipeline 
      (truncated for reading purposes).}
    \label{fig:cluster0}
\end{figure}

\begin{figure}
    \centering
    \includegraphics[width=0.5\textwidth]{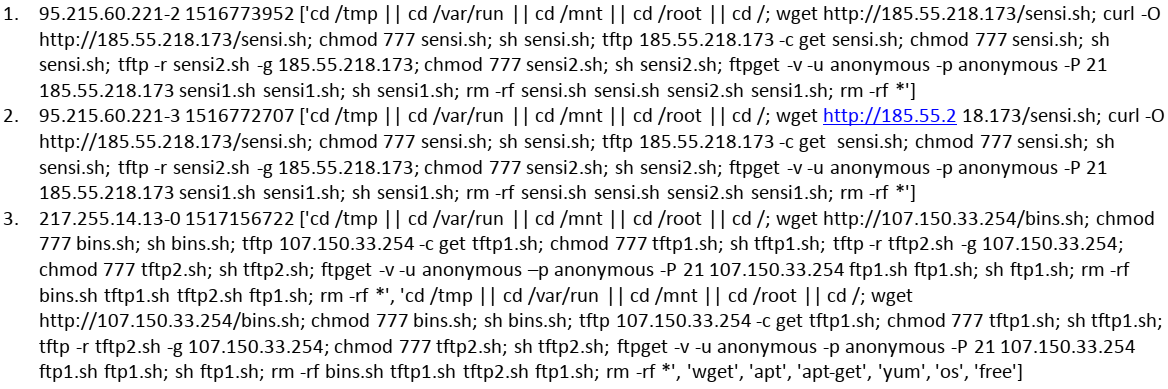}
    \caption{One of the 120 autoencoder session clusters containing
      three sessions with their ID, epoch time stamp, and list of
      commands. To the human eye they appear as variations of a same mode.}
    \label{fig:vdgmm_cluster_0}
\end{figure}

\begin{figure}
    \centering
    \includegraphics[width=0.48\textwidth]{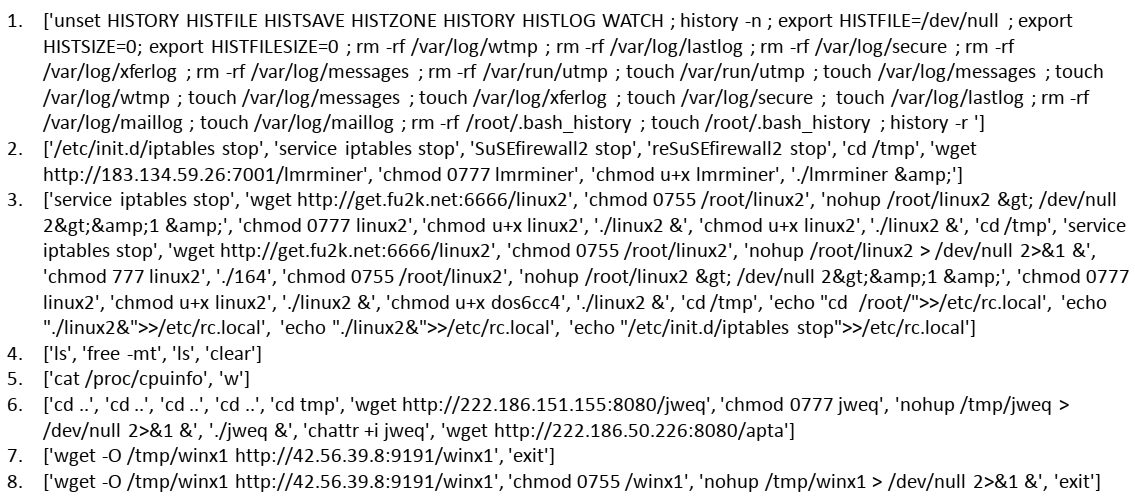}
    \caption{8 representative sessions from as many
      session clusters out of the 120 autoencoder session clusters.}
    \label{fig:vdgmm_clustering}
\end{figure}

The LDA pipeline groups documents according to a predefined number of
clusters (200). A direct inspection of the output shows command list clusters
that are visibly similar to a human eye.
Despite the presence of outliers, the clusters seem quite homogeneous as seen, for
example, in Fig.~\ref{fig:cluster0} which contains a small excerpt of
text from one of the computed clusters. These results indicate that
NLP text processing techniques can be successfully applied to command sequences,
provided that we define an appropriate vocabulary (the tokenization
method we deploy dictates the observable vocabulary features).
The autoencoder pipeline identified 120 different clusters of
sessions.  Figure~\ref{fig:vdgmm_cluster_0} shows one of the clusters
whereas Figure~\ref{fig:vdgmm_clustering} shows an arbitrary selection
of eight different types of sessions by listing the representatives
from eight arbitrarily selected clusters. Clustering at the session
level enables two types of analysis. First, as before, we can recover
botnets and fuzzy signatures. Second, we now have a way to encode a
time series of sessions and from it extract the behavior of each
individual botnet over time, as discussed in \autoref{sect:L2}.



\subsection{L2: analyzing sessions time series.}
\label{sect:L2}

\begin{figure}[ht]
    \centering
    \includegraphics[width=0.5\textwidth]{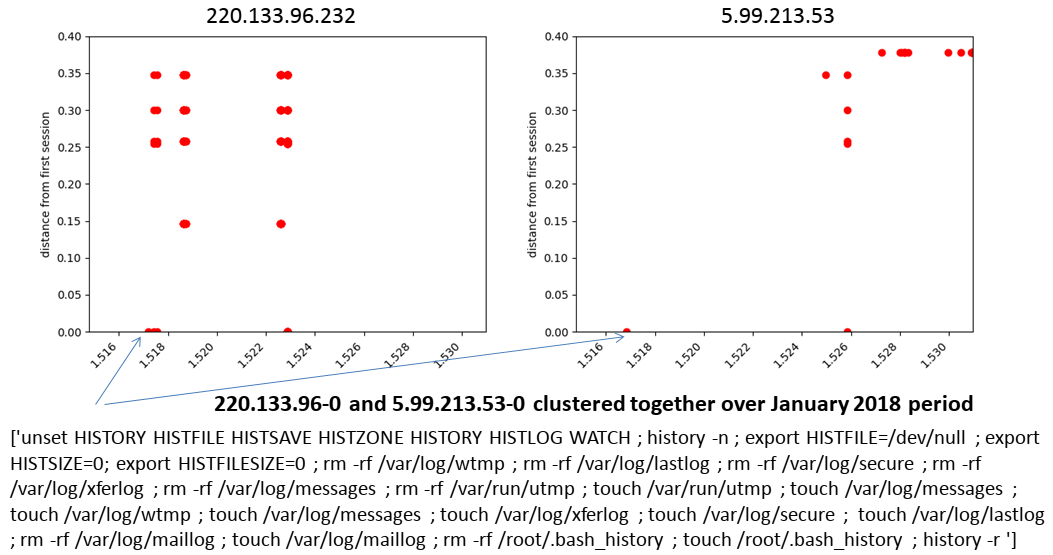}\\
    \vspace{0.2cm}
    \includegraphics[width=0.5\textwidth]{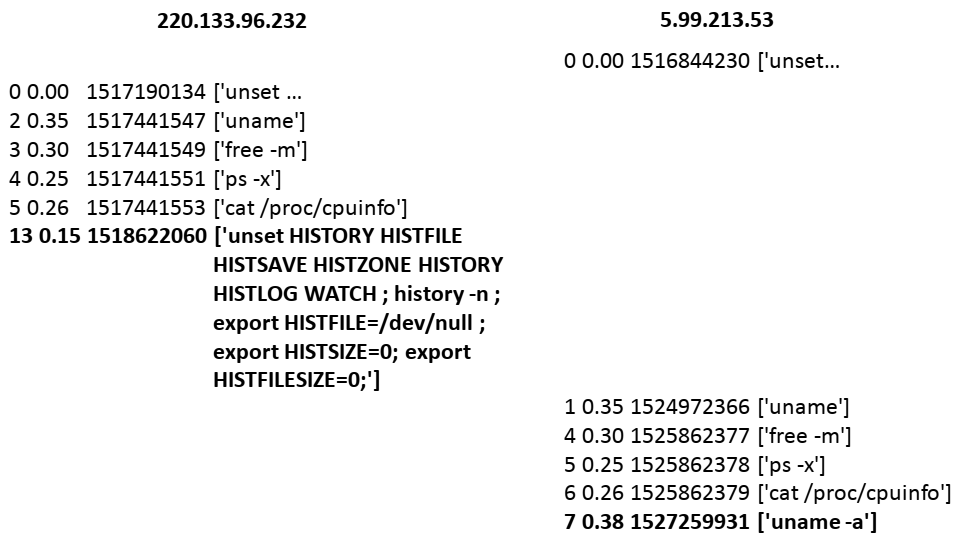}
    \caption{Signals {\tt 220.133.96.132} (left)
      and {\tt 5.99.213.53} (right). Each point $(t,d)$ represents an embedded session $y_t$ that occurred at time $t$ with $d$ the distance from the first signal session. Thus, similar sessions map to points with similar ordinate. While the converse is not necessarily true it is reasonable to hypothesize that the two signals share four fuzzy signatures. Moreover, they seem to exhibit a joint temporal pattern. This intuition is confirmed by the inspection of the corresponding raw sessions (bottom).}
    \label{fig:signals}
\end{figure}

For level L2, we embed each session into a finite
dimensional space where each source IP exhibits a vectored signal
behavior defined as the time series of its encoded sessions: $ip:
y_{t_1},y_{t_2},\ldots,y_{t_n}$, with
$y_{t_i}=E(W(s_i))\in\mathbf{R}^M$. In this short paper we present only preliminary observations about
such signals and defer a more thorough analysis to an upcoming
expanded version of this paper. Figure~\ref{fig:signals} shows the behavior of two signals,
{\tt 20.133.96.132} and {\tt 5.99.213.53}, that start from sessions
that were clustered together by the L1 analysis. For each signal, we plot the distance between its $i^{\rm th}$ session
and its first one recorded. We observe that signal sessions are time
clustered and repetitive; moreover, related bots visibly show
synchronous activities. For example, they are active during
non-overlapping periods of time. These are indications that we can
train probabilistic models capable of capturing the temporal behaviors
of signals (e.g., HSMMs and LSTMs) in order to predict bot's future
actions.


\vspace*{-1em}
\section{Conclusions and Future Work}
\label{sec:conclusions}

We have demonstrated the potential of applying NLP methods to parse
honeypot logs of shell commands and to characterize fuzzy signatures.
We also discussed exploiting session embedding functions to study
high-level, longitudinal botnet behaviors.

We are expanding this work
in three directions:
1) developing fast algorithms to train and
incrementally update word2vec models as suggested
in~\cite{levy:14}; 2) modifying the autoencoder to jointly
train a classification layer as done in~\cite{xie:16}; and 3) building probabilistic models (HMMs, HSMMS, LSTMs) to study
  longitudinal botnet behaviors (time series of encoded sessions) to
  predict future session clusters and to group source IPs exhibiting
  temporally coordinated behaviors.


\bibliography{ml,comp_sec,botnets}

\begin{thebibliography}{10}

\bibitem{andrews:19}
Nicholas Andrews and Marcus Bishop.
\newblock Learning invariant representations of social media users.
\newblock {\em CoRR}, abs/1910.04979, 2019.

\bibitem{Blei05variationalinference}
David~M. Blei and Michael~I. Jordan.
\newblock Variational inference for {D}irichlet process mixtures.
\newblock {\em Bayesian Analysis}, 1:121--144, 2005.

\bibitem{Blei03latentdirichlet}
David~M. Blei, Andrew~Y. Ng, Michael~I. Jordan, and John Lafferty.
\newblock Latent {D}irichlet allocation.
\newblock {\em Journal of Machine Learning Research}, 3:2003, 2003.

\bibitem{chavoshi:17}
Nikan Chavoshi, Hossein Hamooni, and Abdullah Mueen.
\newblock Temporal patterns in bot activities.
\newblock In {\em Proceedings of the 26th International Conference on World
  Wide Web Companion}, pages 1601–--1606, April 2017.

\bibitem{dagon:07}
D.~Dagon, G.~Gu, C.~Lee, and W.~Lee.
\newblock A taxonomy of botnet structures.
\newblock In {\em Proceedings of the Annual Computer Security Applications
  Conference (ACSAC)}, 2007.

\bibitem{guo:17}
Xifeng Guo, Xinwang Liu, En~Zhu, and Jianping Yin.
\newblock Deep clustering with convolutional autoencoders.
\newblock In {\em Liu D., Xie S., Li Y., Zhao D., El-Alfy ES. (eds) Neural
  Information Processing (ICONIP)}, pages 373--382, 10 2017.

\bibitem{rfc4250}
S.~Lehtinen and C.~Lonvick.
\newblock {The Secure Shell (SSH) Protocol Assigned Numbers}.
\newblock RFC 4250 (Proposed Standard), January 2006.

\bibitem{levy:14}
Omer Levy and Yoav Goldberg.
\newblock Neural word embedding as implicit matrix factorization.
\newblock In {\em Proceedings of the 27th International Conference on Neural
  Information Processing Systems (NIPS'14)}, volume~2, pages 2177--–2185,
  December 2014.

\bibitem{kmeans}
James MacQueen.
\newblock Some methods for classification and analysis of multivariate
  observations.
\newblock In {\em Proceedings of the 5th Berkeley Symposium on Mathematical
  Statistics and Probability}, volume 5.1, pages 281--297, 1967.

\bibitem{madiraju2018deep}
Naveen~Sai Madiraju, Seid~M. Sadat, Dimitry Fisher, and Homa Karimabadi.
\newblock Deep temporal clustering: Fully unsupervised learning of time-domain
  features.
\newblock {\em arXiv:1802.01059 cs.LG}, 2018.

\bibitem{provos2007virtual}
N.~Provos and T.~Holz.
\newblock {\em Virtual Honeypots: From Botnet Tracking to Intrusion Detection}.
\newblock Pearson Education, 2007.

\bibitem{stonegross:09}
Brett Stone-Gross, Marco Cova, Lorenzo Cavallaro, Bob Gilbert, Martin
  Szydlowski, Richard~Allen Kemmerer, Christopher Kruegel, and Giovanni Vigna.
\newblock Your botnet is my botnet: analysis of a botnet takeover.
\newblock In {\em Proceedings of the 16th ACM conference on Computer and
  communications security}, pages 635–--647, November 2009.

\bibitem{xie:16}
Junyuan Xie, Ross Girshick, and Ali Farhadi.
\newblock Unsupervised deep embedding for clustering analysis.
\newblock In {\em Proceedings of the 33 rd International Conference on Machine
  Learning, New York, NY, USA}, 2016.

\end{thebibliography}
\bibliographystyle{plain}

\newpage

\renewcommand{\thesection}{\Alph{section}}
\setcounter{page}{1}
\setcounter{section}{0}

\section{Appendix}
\label{appendix}

All honeypot data were downloaded from {\tt
  https://haas.nic.cz/}. Sessions are time-stamped. However,
individual bash commands in each session are not. We tokenized bash
commands with

{\tt
 \begin{verbatim}
def tokenize(line):
    words = re.sub(
            r"(2>&1)|(>&)|(&>)|(>\|)|(>>)|([012]>)",
                    " _r_ ", line)
    words = words.replace(">", " > ")
    words = words.replace("<", " < ")
    words = words.replace("=", " = ")
    words = words.replace("||", " _P_ ")
    words = words.replace("|", " _p_ ")
    words = words.replace("&&", " _A_ ")
    words = words.replace("&", " _a_ ")
    lwords = re.split("[;,\"() ]", words)
    return lwords
\end{verbatim}
}

\subsection{LDA (LSI) pipeline:}

\begin{figure}[h]
    \centering
    \includegraphics[width=0.5\textwidth]{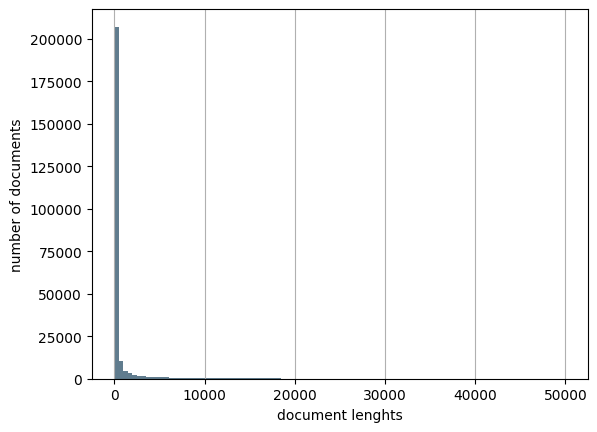}
    \caption{Distribution of document lengths (after removing 3566 documents longer than 50000 tokens).}
    \label{fig:h_exp1}
  \end{figure}

  We parsed sessions collected between 1/2018 and 2/2020 and aggregated them by source IP:
  \begin{center}
\begin{tabular}{ l l } 
  \hline
    Number of tokens: & 331550\\
Source IPs (documents): & 248795\\
Average document length: & 5347.61\\
 \hline
\end{tabular}
  \end{center}
  Figure~\ref{fig:h_exp1} shows the distribution of the lengths of the documents aggregated by IPs where we have removed few but very long documents for ease of display.
  
\subsection{Autoencoder pipeline: full vocabulary:}

We parsed sessions collected between 1/2018 and 10/2020 and treated
them as separate documents; different tokens were mapped to different
indexes:
  \begin{center}
\begin{tabular}{ l l } 
  \hline
  Source IPs: & 420695\\
  Vocabulary size: & 569665\\
  Number of sessions (documents): & 90063468\\
  Average session length: & 15.7\\
 \hline
\end{tabular}
  \end{center}
Figure~\ref{fig:h_mc1} shows the distribution of session lengths.
  
\subsection{Autoencoder pipeline: vocabulary with tokens occurring only once mapped to single word:}

\begin{figure}[h]
    \centering
    \includegraphics[width=0.5\textwidth]{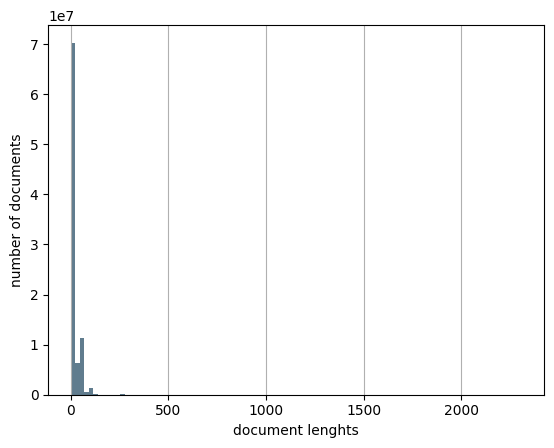}
    \caption{Distribution of session lengths.}
    \label{fig:h_mc1}
  \end{figure}

We parsed sessions collected between 1/2018 and 10/2020 and treated
them as separate documents; tokens occurring only once were mapped to the same index:

\begin{center}
\begin{tabular}{ l l } 
  \hline
Source IPs: & 420695\\
Vocabulary Size: & 426439\\
Number of sessions (documents): & 90063468\\
Average session length: & 15.7\\
\hline
\end{tabular}
\end{center}
Figure~\ref{fig:h_mc1} shows the distribution of session lengths.

\end{document}